# Polarization Profiles of Scattered Emission Lines.
# II. Upstream Dust Scattering in the HH 1 Jet

W. J. Henney
Instituto de Astronomía, UNAM, Apartado Postal 70–264,
04510 México D. F., México

A. C. Raga
Department of Mathematics, UMIST, PO Box 88, Manchester,
M60 1QD, UK

D. J. Axon[1]
Nuffield Radio Astronomy Laboratories, University of
Manchester, Jodrell Bank, Macclesfield, Cheshire, SK11 9DL,
UK

1. Present address: Space Telescope Science Institute, 3700 San Martin Drive, Baltimore, MD 21218, USA

# Abstract

Detailed comparisons are made between observations of scattered light upstream of the head of the HH 1 jet and predictions of simple scattering models. It is shown that, in order to unambiguously determine the velocity of the head of the jet (bow shock) with respect to the upstream dust, existing spectroscopic observations are insufficient and that spectropolarimetric observations of the scattered light are necessary. Such an independent measure of the bow shock velocity is important in order to test "multiple outflow" theories of Herbig-Haro jets. It is also shown that the scattering dust must have a very forward-throwing scattering phase function ($\langle\cos\theta\rangle \sim 0.7$) and slight evidence is found for a dust-gas ratio that is higher than average.

*Subject headings*: ISM: dust, extinction—ISM: jets and outflows—ISM: reflection nebulae—polarization—stars: individual (HH 1)

# 1 Introduction

Herbig-Haro (HH) objects are apparently the result of the interaction of high velocity outflows from young stars (or from protostellar objects) with the surrounding environment (see, e. g., the review of Reipurth 1992, which has a very extensive list of references). It has recently been realized that some of these objects can be quite successfully modelled as bowshocks, which could be formed, e. g., at the head of a jet-like flow ejected by the outflow source (the first detailed HH object bowshock models were discussed by Hartmann & Raymond 1984).

Even though at least partial agreement between bowshock models and observations has been found for many HH objects (see, e. g., Hartigan, Raymond & Hartmann 1987), detailed comparisons between predictions and observations including line profiles, line ratios, and narrow band images have apparently only been carried out for the high excitation objects HH 1 (Raga et al. 1988; Noriega-Crespo, Böhm & Raga 1990; Noriega-Crespo, Böhm & Raga 1989) and HH 34 (Morse et al. 1992), and the low excitation object HH 7 (Curiel 1992). For HH 1 and HH 34, the best agreement with the observed spatially resolved line profiles and line ratios is found for bowshock models with velocities in the $v_{\rm BS} \approx$ 150–200 km s$^{-1}$ range (Raga et al. 1988; Morse et al. 1992). On the other hand, the low excitation object HH 7 is better reproduced with a $v_{\rm BS} \approx 100$ km s$^{-1}$ model (Smith 1991; Curiel 1992)

Another observational constraint is provided by the fact that the observed proper motions and radial velocities allow us to directly determine the true spatial velocity of HH 1 ($v_{\rm S} \approx 352$ km s$^{-1}$; Herbig & Jones 1981) and HH 34 ($v_{\rm S} \approx 320$ km s$^{-1}$; Heathcote & Reipurth 1992; Eislöffel & Mundt 1992). In both cases, the spatial velocities of the sources are likely to be at least an order of magnitude smaller (e. g., no proper motion has been detected for the source of HH 34, and its radial velocity is a factor $\sim 5$ lower than the one of the HH object), so that the $v_{\rm S}$ values also approximately correspond to the velocities of the HH objects relative to the outflow sources. Unfortunately, no proper motion has been detected for HH 7 (Herbig & Jones 1983), so that its spatial velocity is not known. Because of this, in the following discussion we will concentrate on the objects HH 1 and HH 34.

From the results described above, it is evident that for HH 1 and HH 34 a clear discrepancy exists between the $v_{\rm BS} \approx$ 150–200 km s$^{-1}$ velocities deduced from the bowshock models and the $v_{\rm S} \approx$ 300–350 km s$^{-1}$ spatial velocities. There are three plausible explanations for this quite remarkable discrepancy (see, e. g., the discussion of Raga & Kofman 1992) :

1. Due to the difficulty of measuring the positions of the diffuse HH knots, the true errors in the proper motion measurements are much larger than the estimated errors.

2. There is some fundamental problem in the bowshock models (e. g., the geometry assumed for the bowshock is incorrect).

3. The environment of the HH objects is moving radially away from the source at a velocity $v_{\rm E} \approx 150$ km s$^{-1}$.

For the case of HH 1, new determinations of the proper motions (Raga, Barnes & Mateo 1990) clearly confirm the previously determined values (Herbig & Jones 1981), so that, at least for this object, the first explanation appears to be unlikely. Also, as we have described above, the intensity maps and the spatially resolved line profiles and line ratios of HH 1 and HH 34 are well reproduced by bowshock models. Because of this, it would be somewhat surprising if the $v_{\rm BS}/v_{\rm S}$ discrepancy were the result of a fundamental problem with the bowshock models.

We are then left with the third explanation. In this scenario, the bowshocks are moving into a medium that is expanding radially away from the source with a velocity $v_{\rm E} \approx 150$ km s$^{-1}$. In this way, a bowshock with a spatial velocity $v_{\rm S} \approx 350$ km s$^{-1}$ has a relative velocity $v_{\rm BS} \approx 200$ km s$^{-1}$ with respect to the preshock gas, in good agreement with the measurements. However, this explanation has the somewhat undesirable characteristic of introducing a large velocity for the environment as an extra free parameter of the model. If such an environmental velocity indeed existed, it would either imply that the (optically detected) outflow is moving into a slower (undetected and possibly less well collimated) outflow (see, e. g., Stocke et al. 1988), or that it is moving into the slower "tail" of a previous "outflow episode" which was ejected in the past by the same source (see, e. g., Raga et al. 1990).

This conclusion appears to be favored by a new observational result. Solf & Böhm (1991) (hereafter SB) observed faint, highly blue-shifted line profiles in the near environment farther away from the source than HH 1. This is in sharp contrast to the very low radial velocities measured in HH 1 itself (a direct result of the fact that the HH 1/2 outflow axis is almost on the plane of the sky). Noriega-Crespo, Calvet & Böhm (1991) (hereafter NCB) have interpreted these highly blue-shifted line profiles as the result of scattering of the emission of HH 1 by dust present in the near environment of the object. These authors find that in order to reproduce the observed line profiles, it is necessary to assume a relative velocity between HH 1 and its environment of $\approx 200\,\mathrm{km\,s^{-1}}$. This measurement would be completely consistent with the scenario described above (in point number 3).

However, this apparent confirmation of the existence of a moving environment in HH 1 depends on the identification of the blue-shifted line profiles (observed close to HH 1, see SB) as HH 1 emission scattered on environmental dust. Polarization measurements would clearly confirm if this is indeed the case.

In Henney (1994) (Paper I), a general theoretical framework was constructed for the interpretation of spectropolarimetric observations of dust-scattered emission lines. Predictions were presented of the spatially- and velocity-resolved intensity and polarization that would result from the Rayleigh scattering of the emission of a moving object from surrounding dust. The details of the extension of these models to non-Rayleigh scattering from larger dust grains will be presented in Henney, Axon & Raga (1994) (Paper III). In this paper, such models are fitted in detail to observations of the upstream light in the HH 1 jet in an attempt to extract information about the physical parameters of the jet and its surroundings.

## 2 Scattering Models for HH 1

### 2.1 Integrated Scattered Light

Solf & Böhm (1991) (SB) present a contour image of the HH 1/2 complex in the light of [S II]$\lambda\lambda$ 6716, 6730 (their Figure 1). From this image, it is possible to *very crudely* estimate the scattering optical depth of the surrounding dust cloud in the following fashion. If it is assumed that the lowest two contours in the region around and upstream of the leading condensation HH 1F are entirely due to scattered light and that all the higher contours are due to intrinsic emission, then, by measuring the area under each contour, it can be calculated that the ratio of scattered to intrinsic flux is $\sim 0.2$. For a spherical cloud, as is shown in Paper I, this ratio is equal to the scattering optical depth to the source, so

$$\tau_{\mathrm{scat}} = \sigma^{\mathrm{H}} n_{\mathrm{H}} (R_c - R_0) \simeq 0.17, \qquad (1)$$

where $n_{\mathrm{H}}$ is the hydrogen number density, $\sigma^{\mathrm{H}}$ is the mean scattering cross-section per *hydrogen atom* at the appropriate wavelength and $R_c$, $R_0$ are the radii of the scattering cloud and the source. The value of $\sigma^{\mathrm{H}}$ depends on the dust-gas ratio and on the composition and size distribution of the dust. Possible variations in this parameter are discussed in Section 2.2 below but initially a value of $4.6 \times 10^{-22}\,\mathrm{cm}^2$ is adopted, as by NCB, to best allow comparison with their results. For the same reason, a value for the distance to HH 1 of 500 pc, and for the radius of the source (working surface) of $3''$ are used, giving

$$(R_{15} - 0.2) n_{750} \simeq 4.4, \qquad (2)$$

where $R_{15}$ is the scattering cloud radius in units of 15" and $n_{750}$ is the cloud's hydrogen number density in units of $750\,\mathrm{cm}^{-3}$. The radius and density are expressed in these units because they are the values derived by NCB in fitting their radiative transfer model to the longslit spectrum of SB. Their alternative model has $R_c = 28.5''$ and $n_{\mathrm{H}} = 500\,\mathrm{cm}^{-3}$, so they find the right-hand-side of the above equation to be 0.8 or 1.13. The sole reason for the discrepancy between these values and that of the current work is that NCB estimate a much smaller optical depth ($\tau_{\mathrm{scat}} = 0.03$–$0.04$) and it is therefore worth investigating the validity of the assumptions behind the derivation of equation (1).

Inaccuracies can arise either because of errors in determining the ratio of scattered to intrinsic flux or because the geometry and physical state of the scattering cloud render equation (1) inappropriate. Considering the flux ratio first, it is unlikely that a significant proportion of the light in the lowest two contours of SB's image is due to intrinsic emission. This is because the [S II] longslit spectrogram along p.a. 329° shows that the faint upstream emission is kinematically distinct from both the emission from the working surface of the jet and that from the foreground nebula (unlike in the case of H$\alpha$, where there is significant nebular contamination). Much more likely is that a significant amount of the scattered light falls in the region of the higher contours and is thus swamped by the intrinsic emission. This is expected on theoretical grounds since the models suggest that the intensity of scattered light should be strongly peaked near the source, especially for forward-throwing scattering (see Paper III), and evidence for it can be seen in the broad wings of the line profile from the working surface. Hence, the amount of scattered light is probably underestimated rather than overestimated, so equation (1) is a *lower limit* on $\tau_{\rm scat}$.

Considering next the idealizations behind equation (1), it is possible that extinction should be included since, with a mean grain albedo $\omega \sim 0.5$ at this wavelength (Draine & Lee 1984), the extinction optical depth would be $\sim 0.4$. The apparent intensity of the source $I_0'$ is then given by

$$I_0' = I_0 \exp\{-\tau_{\rm ext}\} = I_0 \exp\{-\tau_{\rm scat}/\omega\}, \qquad (3)$$

since any light absorbed or scattered will no longer appear to come from the source. The scattered light, however, will only be attenuated by absorption since any further scattering out of the beam will be exactly compensated by scattering into it. On the other hand, the path travelled in the cloud by a photon which has been scattered once will be longer than that of an unscattered photon by a factor

$$\beta = \bar{R}(1+\cos\theta) + (1-\bar{R}^2\sin^2\theta)^{1/2}, \qquad (4)$$

where $\theta$ is the scattering angle and $\bar{R}$ is the radius at which the photon was scattered, in units of $R_c$. The mean value of $\beta$ for the scattered light [1] can then be calculated as

$$\langle\beta\rangle = \frac{3}{2}\int_0^1\int_0^\pi \beta X(\theta)\bar{R}^2\sin\theta\,d\theta\,d\bar{R}, \qquad (5)$$

where $X(\theta)$ is the scattering phase function. For small dust grains, this will be the Rayleigh phase function (see Paper I), and, for larger grains, a modification (West 1991) of the empirical Henyey-Greenstein (H-G) phase function (Henyey & Greenstein 1941). This latter function has one free parameter $\tilde{g}$, which describes the asymmetry of the scattering and is related to the mean cosine of the scattered light via

$$\langle\cos\theta\rangle = \tilde{g}\frac{3(4+\tilde{g}^2)}{5(2+\tilde{g}^2)}. \qquad (6)$$

It is preferable to use an empirical phase function, rather than one derived from simple physical theory, when considering the angular dependence of scattering from large dust grains because the scattering properties are significantly affected by the porosity and surface roughness of the grains, which cannot easily be modelled in a straightforward fashion. This topic will be discussed further in Paper III of this series. The quantity $\langle\beta\rangle$ is plotted in

---

[1] Since this is only a first order correction to the scattered intensity, it is not necessary to consider multiply-scattered photons when calculating this path.

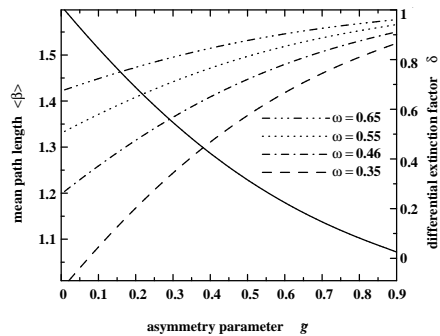

Figure 1: Differential extinction of source and scattered light is a result of two competing factors: a singly scattered photon will, on average, have travelled a distance in the dust cloud longer than that of an unscattered photon by a factor $\langle\beta\rangle$ (solid line) but, on the other hand, only absorption need be considered in calculating the attenuation of scattered light (since out- and in-scattering will balance) whereas the source light is attenuated by absorption *and* scattering. The resulting differential extinction factor (dashed and dotted lines) is shown for a range of albedos $\omega$ as a function of the asymmetry parameter of the scattering phase function.

Figure 1 as a function of scattering asymmetry parameter $\tilde{g}$ (solid line) for the modified H-G phase function (*cf.* West 1991 and Paper III). The apparent intensity of the scattered light $I_s'$ is then related to the scattered intensity if there were no extinction $I_s$ by

$$I_s' = I_s \exp\{-(1-\omega)\langle\beta\rangle\tau_{\rm scat}/\omega\}, \qquad (7)$$

and so the ratio of apparent scattered intensity to source intensity is

$$\frac{I_s'}{I_0'} = \frac{I_s}{I_0}\exp\{-[(1-\omega)\langle\beta\rangle-1]\tau_{\rm scat}/\omega\} = \tau_{\rm scat}\exp\{\delta\tau_{\rm scat}\}, \qquad (8)$$

where a relative extinction factor

$$\delta = [1-(1-\omega)\langle\beta\rangle]/\omega \qquad (9)$$

has been introduced. This relative extinction factor is also plotted as a function of scattering asymmetry parameter in Figure 1, for various values of the mean grain albedo $\omega$. As can be seen, if the albedo and the scattering asymmetry are both small, then $\delta$ is negligible and the source and scattered light are both attenuated to the same degree, so the extinction can be safely ignored. Even with relatively high albedos and a very forward-throwing phase function, the extinction factor is always less than unity, as can be seen by considering the limit of equation (9) as $\langle\beta\rangle$ tends to one. In this worst case of $\delta = 1$, then, using equation (8), a new estimate for the scattering optical depth of $\tau_{\rm scat} \sim 0.15$ can be made. Notice that this is not significantly smaller than the original estimate.

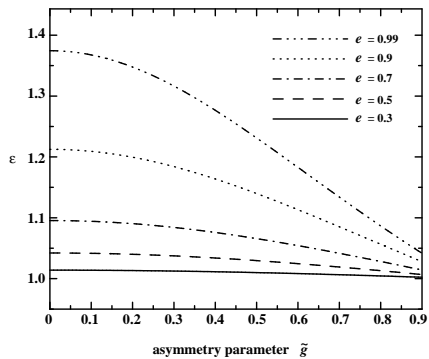

Figure 2: Boost factor $\epsilon$ for scattered intensity from a prolate ellipsoidal scattering cloud over that which would be seen from a spherical cloud having the same optical depth to the source. The major axis of the cloud lies in the plane of the sky and $\epsilon$ is plotted against the asymmetry factor $\tilde{g}$ of the scattering phase function (modified H-G) for a variety of cloud eccentricities.

Turning now to the geometry of the scattering cloud, equation (1) can be extended to non-spherical clouds by replacing $\tau_{\rm scat}$ (the optical depth along the line of sight to the source) with a mean optical depth, averaged over all scattered photons, which is given by

$$\langle \tau_{\rm scat} \rangle = \frac{\tau_{\rm scat}}{4\pi \, R_{\rm c}|_{\rm l.o.s.}} \int_0^{2\pi} \int_0^{\pi} R_{\rm c}(\theta, \phi) X(\theta) \sin\theta \, d\theta \, d\phi$$
$$\equiv \epsilon \tau_{\rm scat}, \qquad (10)$$

where $R_{\rm c}|_{\rm l.o.s.}$ is the cloud radius along the line of sight to the source and $\epsilon$ is the ratio of scattered intensity to that which would be seen from a spherical cloud having the same scattering optical depth to the source. For the simple case of a prolate ellipsoid of eccentricity $e$ whose major axis lies in the plane of the sky, then $\epsilon$ is given by

$$\epsilon = \frac{1}{4\pi} \int_0^{2\pi} \int_0^{\pi} \frac{X(\theta) \sin\theta}{(1 - e^2 \cos^2\theta)^{1/2}} d\theta \, d\phi. \qquad (11)$$

This is plotted in Figure 2 as a function of scattering asymmetry parameter for eccentricities of 0.3–0.99 and it can be seen that, even for highly eccentric ellipsoids ($e = 0.99$ corresponds to a major:minor axis ratio $a/b \simeq 7$), the enhancement to the scattered intensity is not large, especially for high values of $\tilde{g}$. Hence, it seems that, even with a very non-spherical cloud, equation (1) will be in error by less than a factor of two. It must be remembered, however, that $R_{\rm c}$ in that equation is the cloud radius along the line of sight to the source, which is the *minor* axis of the ellipse in the example considered above. The major axis of the ellipse (which will determine the maximum extent of scattered light upstream of the source) could, depending on the eccentricity, be much greater. A non-spherical cloud would also affect the value of $\langle \beta \rangle$ but this will not be calculated here.

Combining the above considerations and adopting $\tilde{g} = 0.6$, $\omega = 0.55$ and $e = 0.9$ (giving $\delta = 0.85$, $\epsilon = 1.1$) leads to a new lower bound on the scattering optical depth to the source of $\tau_{\rm scat} > 0.13$. While this is lower than the original estimate, it is still much higher than the value obtained by NCB. Obviously, there are further ways in which the analysis could be refined. For instance, none of the above calculations take account of the finite size of the source, despite the fact that this is implicit in equation (2). However, the crudity of the determination of the scattered light fraction means that further refinement would be unwarranted.

The *extinction* optical depth to HH 1 can be determined independently, from the observations of Solf, Böhm & Raga (1988), who measure a reddening $E[B - V] = 0.43$ from which (assuming a total-to-selective extinction ratio $R_V = 3.1$; Mathis 1990) it follows that $\tau_{\rm ext} = 1.22$. If all the reddening is local to HH 1, then (with $\omega = 0.55$) this implies $\tau_{\rm scat} = 0.67$ but, since the reddening could occur anywhere along the line of sight, this is an *upper* limit to the scattering optical depth and is hence quite consistent with the value derived above.

## 2.2 Spatial Distribution of the Scattered Light

The [SII] spectrogram of SB is taken with the slit lying along the HH 1 jet and so corresponds to Aperture A described in Paper I. It is possible to estimate the velocity-integrated scattered intensity as a function of distance along this slit. The slope of the scattered brightness profile measured in this manner is an independent observation from that of the total scattered intensity and should, in principle, allow the radius of the scattering cloud to be determined. Unfortunately, the situation is rather less straightforward than this and the derived cloud radius is strongly dependent on the assumed cloud shape and scattering phase function. In Figure 3, three different models are presented, all of which fit the brightness profile equally well.

Model 1 is for a spherical cloud with Rayleigh scattering. The surface brightness profile is plotted for scattering from a finite source (solid line) and a point source (dot-dashed line). In each case, the source profile of a thin spherical shell (of radius 3″) has been added (dotted line) and the scattering optical depth to the source has been taken to be 0.13. The filled circles signify the measured intensity at 2″ intervals along the slit and the horizontal scale has been adjusted to give the best fit between model and observations. The degree of polarization is also shown. It can be seen that the scattered intensity hardly differs between the point and extended source, although the polarization is lower for the extended source. The cloud radius deduced from the fit is $R_{\rm c}=18''$, which, using $\tau_{\rm scat}=0.13$, implies a number density for the upstream gas of $n_{\rm H} \simeq 3300\,{\rm cm}^{-3}$.

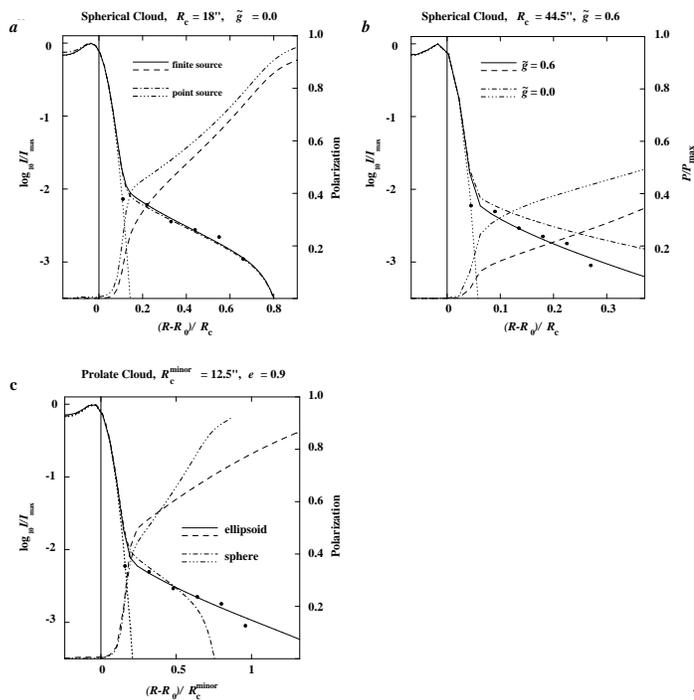

Figure 3: Three models for the brightness profile in Aperture A. Measurements from the [SII] spectrogram of SB at 2″ intervals are shown as filled circles. The edge of the source is marked by a vertical line and the source profile (smeared by seeing of 1″) by the dotted line. a. Model 1. The solid line shows the profile of the model (source plus scattered light) and the dot-dashed line shows the result of calculating the scattering using a point source. The dashed and dot-dot-dot-dashed lines show the degree of polarization for these two cases. b. Model 2 (non-Rayleigh scattering). The solid line shows the profile of the model (source plus scattered light), which uses a scattering asymmetry parameter $\tilde{g}=0.6$, and the dot-dashed line shows the profile of Model 1 ($\tilde{g}=0$) to the same scale. The dashed and dot-dot-dot-dashed lines show the degree of polarization for these two cases, normalized to the maximum polarization for that phase function. c. Model 3 (ellipsoidal cloud). The solid line shows the profile of the model (source plus scattered light), in which the scattering cloud has an eccentricity $e=0.9$, and the dot-dashed line shows the profile of Model 1 to the same scale. The dashed and dot-dot-dot-dashed lines show the degree of polarization for these two cases.

TABLE 1
Scattering Models that Fit the Observed Brightness Profile

| Parameter | Model 1 | Model 2 | Model 3 |
|---|---|---|---|
| $\tau_{\rm scat}$ | 0.13 | 0.13 | 0.13 |
| $\tilde{g}$ | 0.0 | 0.6 | 0.0 |
| $e$ | 0.0 | 0.0 | 0.9 |
| $R_{\rm c}$ (″) | 18 | 44.5 | 12.5, 28.5 |
| $n_{\rm H}$ (cm$^{-3}$) | 3300 | 1200 | 4000 |
| $p$ | 0.35–0.7 | 0.05–0.15 | 0.55–0.75 |

Model 2 is also for a spherical cloud but this time using a modified H-G phase function with $\tilde{g}=0.6$. Again, $\tau_{\rm scat}$ is taken to be 0.13 and the Rayleigh scattering model with the same parameters is shown for comparison (dot-dashed line). The forward-throwing scattering causes the intensity to fall off more sharply with distance than in the Rayleigh case and so the horizontal scale needs to be stretched to agree with the observations, leading to a cloud radius $R_{\rm c}=44.5''$ and a number density $n_{\rm H}=1200\,{\rm cm}^{-3}$.

Model 3 reverts to a Rayleigh phase function but uses an ellipsoidal cloud (the major axis of which lies in the plane of the sky) of eccentricity $e=0.9$ ($a/b \simeq 2.3$). As usual, $\tau_{\rm scat}=0.13$ and a spherical cloud with radius equal to the minor axis of the ellipsoid is plotted for comparison (dot-dashed line). In contrast to Model 2, the intensity now falls off less steeply with distance (and extends out beyond $R_{\rm c}^{\rm minor}$) and so the horizontal scale needs to be compressed to fit the observations, leading to $R_{\rm c}^{\rm minor}=12.5''$, $R_{\rm c}^{\rm major}=28.5''$ and $n_{\rm H}=4000\,{\rm cm}^{-3}$.

The important features of the three models are summarized in Table 1. Also included in the table is a range of polarizations predicted for the scattered light. The first value is for 4″ upstream of the source and the second value is for 12″ upstream. For non-Rayleigh scattering (Model 2), a Rayleigh form is still used for the polarization phase function but with the maximum degree of polarization $p_{\text{max}}$ (at a scattering angle of 90°) allowed to be less than unity. This is a reasonable approximation to experimental results for scattering from irregularly shaped particles (see Giese et al. 1978 and discussion in Paper III). The value of $p_{\text{max}}$ appropriate to the phase function used in Model 2 is $\sim 0.5$.

Because it is the *slope* of the brightness profile that is being fitted here, the vertical positioning of the observational points as a whole with respect to the model curve has been allowed to vary when finding the best fit. However, for all three models it turns out that the ratio of scattered to source intensity in the slit is within 20% of that measured. Considering the uncertainties in this measurement, this is strong confirming evidence that the scattering optical depth adopted for the models, determined from the total scattered intensity, is essentially correct. As can be seen, the sizes, and hence densities, of the models vary considerably. Model 2 has the lowest density because its strongly forward-throwing scattering phase function requires a very large cloud to reproduce the observed brightness profile. Model 3, on the other hand, has to have a high density because its ellipsoidal shape means that it extends much further in the direction of the slit than perpendicular to it and is hence smaller than a spheroidal cloud would be with a similar slit profile. Note that the predicted polarizations, also, have a wide spread between the models. Model 3 has the largest polarization because its elongated shape leads to the scatterers being concentrated towards the plane of the sky, with scattering angles nearest 90°. Model 2 has the lowest polarization because, first, the high $\tilde{g}$ means that most of the scattered light is scattered through small angles and, second, the peak polarization for this phase function is only 0.5.

Since it is quite possible that asymmetric scattering *and* an aspherical scattering cloud are present in HH 1, then it could be argued that the extremes of Models 2 and 3 will cancel each other out, leaving the parameters derived from Model 1, despite the naivete of its assumptions, as the most appropriate. However, since it is the dust with a projected position near the source that is most important for forward-throwing scattering, the eccentricity of the cloud is relatively unimportant for high values of $\tilde{g}$, as can be seen from Figure 2, and an ellipsoidal cloud would behave in a very similar fashion to a spherical cloud of radius equal to its semiminor axis. Hence, Model 2 should be roughly appropriate to non-Rayleigh scattering, whatever the geometry of the cloud.

Although Models 1, 2, and 3 all fit the observed distribution of scattered light quite well, they also predict embarrassingly large densities for the scattering region, especially Models 1 and 3. Although line ratio diagnostics (Solf, Böhm & Raga 1988) imply electron densities of $\sim 3000$–$10000 \, \text{cm}^{-3}$ at the bow shock (where the gas will be fully ionized), this is gas that has just passed through a radiative shock and the pre-shock gas should be at a much lower density. The line ratios upstream of the bow shock are no guide to the density there, since most of the light is scattered, but Hartigan (1989) estimates $n_{\text{H}} \sim 200$–$400 \, \text{cm}^{-3}$, using general arguments from the kinematics and dynamics of the jet. This is much lower than that derived from the models, which suggests either that the material directly upstream of the bow shock is at a lower density than is typical of its vicinity or that too low a value of $\sigma^{\text{H}}$ has been used. The latter option is quite possible since values of this quantity derived for HII regions can vary by an order of magnitude (Osterbrock 1974). The first option, on the other hand, could pertain if HH 1 was moving into a channel that had been swept out by previous episodes of jet activity (see the discussion of time-varying jets in the introduction).

### 2.3 Spectral Line Shapes of the Scattered Light

The peak velocity and velocity width (FWHM) of the scattered [SII] light, as a function of position along the slit, can be found from Figure 5 of SB. It is also possible to estimate the skewness of the line shapes by combining this information with that contained in the longslit spectrogram. If the half-width-half-maxima to the blue and to the red of the peak velocity are denoted by $\Delta v_{\text{blue}}$ and $\Delta v_{\text{red}}$ respectively, then a simple estimate of the skewness is

$$\kappa = \frac{\Delta v_{\text{blue}} - \Delta v_{\text{red}}}{\Delta v_{\text{blue}} + \Delta v_{\text{red}}}. \quad (12)$$

These quantities ($v_{\text{peak}}$, $\Delta v_{\text{FWHM}}$, and $\kappa$) are shown in Table 2 for various positions along the slit.

In Figure 4 these observational data are compared with predictions from the scattering models that successfully reproduced the spatial distribution of scattered light in the previous section. There are now two further relevant parameters for the models: the source velocity magnitude $u_0$ and the inclination of this velocity to the plane of the sky $\alpha$. Since the intrinsic light from the bow shock of HH 1 has an extremely low radial velocity, it is obvious that the motion of the HH 1 jet is very close to the plane of the sky. Hence, to start with, $\alpha = 0$ will be assumed. The following points should be borne in mind when interpreting these results:

1. The observational velocities are all heliocentric, whereas the model velocities assume the scatterers have no velocity component towards the observer.

TABLE 2
Lineshape Parameters of the [Sii]$\lambda\lambda 6716$, 6730Å Emission Lines

| $R - R_0$ (″) | $v_{\rm peak}$ (km s$^{-1}$) | $\Delta v_{\rm FWHM}$ (km s$^{-1}$) | $\kappa$ |
|---|---|---|---|
| 0 | 0±9 | 79±20 | ⋯ |
| 2 | -36±9 | 113±25 | 0.46±0.12 |
| 4 | -41±9 | 150±25 | 0.60±0.07 |
| 6 | -74±6 | 167±25 | 0.24±0.10 |
| 8 | -94±16 | 148±40 | 0.18±0.20 |
| 10 | -119±19 | 202±40 | 0.04±0.12 |

NOTE.—Measured at various distances upstream from the edge of the leading condensation in HH 1. All values were measured from the data of SB. The errors are estimated from the "noisiness" of the data and do not take into account any contamination of the scattered light by intrinsic emission.

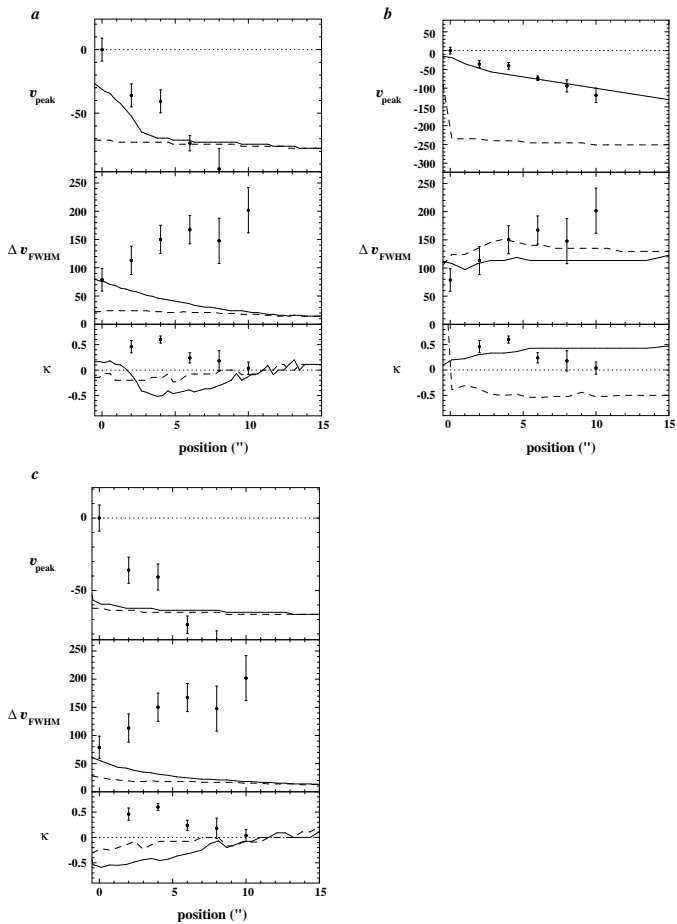

Figure 4: The velocity moments of the total scattered intensity (solid line) and the polarized intensity (dashed line) as a function of upstream distance from the source for the three models of Table 1. In each case, $\alpha = 0$ is assumed and the source velocity magnitude $u_0$ is varied so as to provide the best fit to $v_{\rm peak}$. The filled circles show the observational results (total intensity) of SB. **a**. Model 1, Rayleigh scattering, $R_c = 18''$, $u_0 = 80\,{\rm km\,s^{-1}}$. **b**. Model 2, Modified Henyey-Greenstein scattering, $\tilde{g} = 0.6$, $R_c = 44.5''$, $u_0 = 270\,{\rm km\,s^{-1}}$. **c**. Model 3, Rayleigh scattering, ellipsoidal dust cloud, $e = 0.9$, semimajor axis 28.5″, $u_0 = 70$ km s$^{-1}$.

This is probably unimportant since the line-of-sight velocity of the molecular cloud associated with HH 1 is very low (Rodriguez 1989).

2. The error bars on the observational points were estimated from the noisiness of the data at each position. They are much lower than the formal velocity resolution of the spectrograph (55 km s$^{-1}$).

3. The observational velocities were determined by fitting Gaussians to the line profiles, whereas, with the model velocities, the maximum of each line profile was determined. For a skew line profile, the difference between the velocities deduced by these two methods could be of order $\sim 0.5\kappa\Delta v_{\rm FWHM}$.

Also shown in the figures are the corresponding quantities for the polarized intensity. For each of Models 1, 2, and 3, $u_0$ is varied so as to find the best fit between the observations and model predictions for $v_{\rm peak}$. It can be seen that, for Models 1 and 3, the fit is so poor that the choice of $u_0$ is quite arbitrary. The poor fit is because the observations show a gradual increase in the blue shift of the light as the distance from the source is increased, whereas Models 1 and 3 show a roughly constant blue shift of $-u_0$ for all upstream positions (except for very close to the source in Model 1). This can also be seen in the relevant position-velocity diagram (left panel of Figure 5), where the scattered intensity shows a ridge along $v = -1$ for $\bar{x}$ greater than the source radius (zero in the case of Model 3). It is also clear from the further position-velocity diagrams in Figure 13 of Paper I that varying $\alpha$ will not improve the quality of the fits, since the ridge still exists for $\alpha = 0.5$.

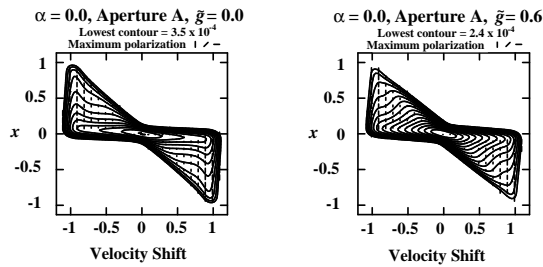

Figure 5: Position-velocity diagrams of scattered light corresponding to two different models for the dust scattering phase function. LEFT PANEL (from Figure 13 of Paper I): Rayleigh scattering (Model 1). Note that the peak intensity of the scattered light is at $v = -1$ (corresponding to the source velocity) for all upstream positions. RIGHT PANEL: Non-Rayleigh scattering (Model 2). Note that the peak intensity of the scattered light is increasingly blue-shifted with increasing distance upstream of the source.

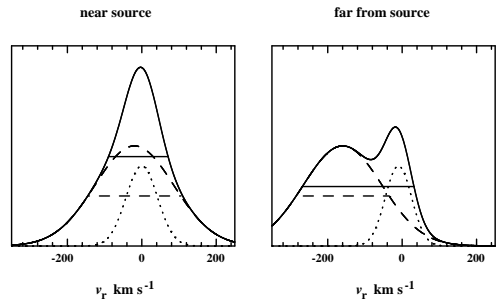

Figure 6: Schematic illustration of how contamination by intrinsic emission can affect the measured line widths. The scattered component is shown by the dashed line and the intrinsic component by the dotted line. The width of the total profile (solid line) is smaller near the source than it is far away from it, despite the fact that the width of the scattered component does not change.

Model 2, on the other hand, fits the observed velocities much better. This is because the asymmetry of the scattering causes the majority of the scattered light to come from dust on the near side of the cloud, which sees an increasingly blue-shifted source as the distance upstream is increased, whereas, with Rayleigh scattering, it is the dust *directly* upstream of the source which contributes most to the scattered intensity from a given line of sight and this dust sees the full blue shift. The phenomenon can also clearly be seen in the position-velocity diagram (right panel of Figure 5), where a ridge in the scattered intensity roughly follows the line $v = -\bar{x}$. Even Model 2, though, is not a perfect fit to the observations of $v_{\rm peak}$, particularly near the source, where the observations suggest $v_{\rm peak} \to 0$. This, however, may be due to contamination by intrinsic emission.

If the velocity width $\Delta v_{\rm FWHM}$ is now considered, it can be seen that, again, Models 1 and 3 are very bad fits to the observations. They both show $\Delta v_{\rm FWHM}$ decreasing sharply with distance, whereas the observations show it increasing. Model 2 shows a roughly constant $\Delta v_{\rm FWHM}$, which also appears to be inconsistent with the observations. However, as illustrated in Figure 6, contamination by intrinsic emission would lead to $\Delta v_{\rm FWHM}$ being underestimated near the source and yet *overestimated* far from the source. This apparent paradox arises in the following manner. The scattered and intrinsic light in the vicinity of the source have similar velocities but the scattered light is much broader so that, even if the intrinsic light is fainter than the scattered light, the FWHM of the sum of the two components is essentially that of the intrinsic light. Further upstream, however, the intrinsic light, which is probably from a foreground nebula and which can be clearly seen in the H$\alpha$ spectrogram of SB, is much less blue shifted than the scattered light so that the FWHM of the blend of the two components is larger than that of either of them individually. If this hypothesis were true, then it may also affect the peak velocity of the light at large distances. There is slight evidence that $v_{\rm peak}$ starts to increase again beyond $\sim 12''$ from the source but the data here are very noisy. There are also faint narrow knots of emission visible near zero velocity from 16 to 26$''$ upstream in the [S II] spectrogram.

Turning now to the skewness $\kappa$ of the line profiles, it can be seen that, yet again, Model 2 agrees, at least qualitatively, with the observational results, while Models 1 and 3 do not even produce the correct sign. Since the skewness is, in effect, the third velocity moment of the specific intensity and was very crudely determined from the data, it is heartening that the agreement with Model 2 is as good as it is, especially since $\kappa$ will also be very sensitive to intrinsic contamination (which perhaps explains the decline at large distances from the source).

The effects of extinction on the line profile of the total scattered light from the cloud are similar in some ways to those of asymmetric scattering, as will be demonstrated in Paper III, so it might be thought that the inclusion of extinction would allow Model 1 to fit the observations better. This is not the case, however, since even using an extinction optical depth $\tau_{\rm ext} = 1.0$ does not significantly change $v_{\rm peak}$ of the upstream scattered light. This can

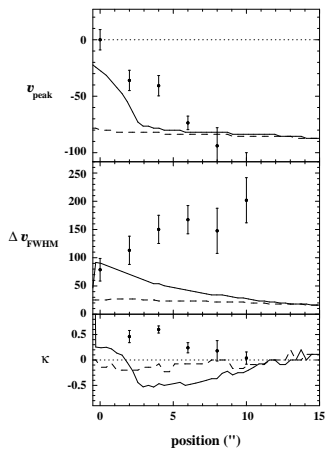

Figure 7: The same as Figure 4 but for Model 1a, which differs from Model 1 in that extinction has been included with an optical depth $\tau_{\text{ext}} = 1.0$. The source velocity magnitude $u_0 = 90\,\text{km}\,\text{s}^{-1}$ and $\alpha = 0$. The filled circles show the observational results (total intensity) of SB and it can be seen that even such a large optical depth does not improve the fit significantly.

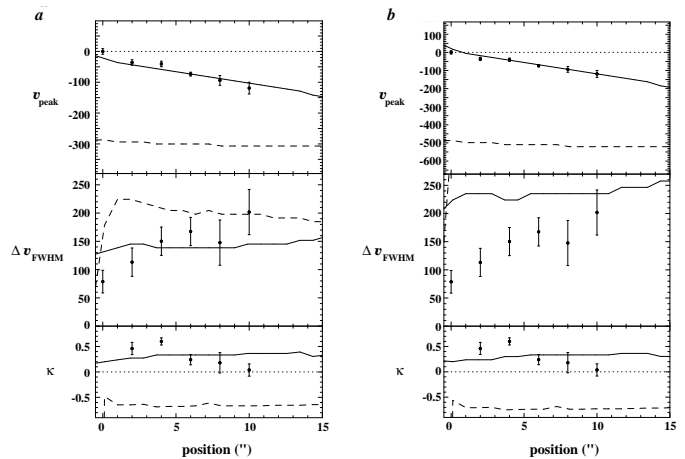

Figure 8: The same as Figure 4 but for Models 2c and 2d, which differ from Model 2 in that they have different inclinations of the source velocity to the plane of the sky $\alpha$. **a**. Model 2c, $\alpha = -0.1\,\text{rad} \simeq -6°$, $u_0 = 330\,\text{km}\,\text{s}^{-1}$. **b**. Model 2d, $\alpha = -0.2\,\text{rad} \simeq -12°$, $u_0 = 560\,\text{km}\,\text{s}^{-1}$. The filled circles show the observational results (total intensity) of SB.

TABLE 3
Source Velocity $u_0$ from Scattering Models

| Model | $\tilde{g}$ | $e$ | $\tau_{\text{ext}}$ | $\alpha$ (°) | $u_0$ ($\text{km}\,\text{s}^{-1}$) | RMS dev |
|---|---|---|---|---|---|---|
| 1 | 0.0 | 0.0 | 0.0 | 0.0 | 80 | 28 |
| 1a | 0.0 | 0.0 | 1.0 | 0.0 | 90 | 26 |
| 2 | 0.6 | 0.0 | 0.0 | 0.0 | 270 | 15 |
| 2a | 0.6 | 0.0 | 0.0 | 5.7 | 190 | 22 |
| 2b | 0.6 | 0.0 | 0.0 | 11.5 | 150 | 27 |
| 2c | 0.6 | 0.0 | 0.0 | -5.7 | 330 | 14 |
| 2d | 0.6 | 0.0 | 0.0 | -11.5 | 560 | 10 |
| 3 | 0.0 | 0.9 | 0.0 | 0.0 | 70 | 38 |

Note.—Derived from an unweighted least-squares fit to the observed values of the blue shift of the scattered light $v_{\text{peak}}$. Also shown is the RMS deviation of the observational data from this fit.

be seen in Figure 7, in which the velocity moments are plotted for such a model (Model 1a).

Accepting that $\tilde{g} = 0.6$ provides the best fit to the observations, the effects of varying $\alpha$ will now be investigated, to see if the fit can be improved. Table 3 summarizes the parameters and quality of fit for all the models considered so far, together with four further models (2a–2d), which are identical to Model 2 except for different values of $\alpha$. It can be seen from the table that Models 2a and 2b (positive $\alpha$) fit the data less well than Model 2. Models 2c and 2d (negative $\alpha$), on the other hand, seem, at first glance, to fit the data better and are shown in Figure 8. However, the source velocity deduced from the fit of Model 2d is $560\,\text{km}\,\text{s}^{-1}$, which is far too high. Raga, Barnes & Mateo (1990) found a velocity for HH1 of $350\,\text{km}\,\text{s}^{-1}$ from proper motion measurements and it seems highly unlikely that $u_0$ can be higher than this. This model would also predict that the direct light from the source should be redshifted by $u_0 \sin 0.2 \simeq 100\,\text{km}\,\text{s}^{-1}$, which is not observed, and produces values of $\Delta v_{\text{FWHM}}$ much higher than those seen. Model 2c predicts $u_0 = 330\,\text{km}\,\text{s}^{-1}$, which is possible if the upstream dust is stationary, but this model is also rather inconsistent with the observed direct light from the source, which suggests a bow shock moving in the plane of the sky. Model 2c does, however, produce the best fit to the observed $\Delta v_{\text{FWHM}}$, although this is quite sensitive to the assumed velocity width of the source. If a higher value had been used (perhaps more realistically) then Model 2 would fit the observed $\Delta v_{\text{FWHM}}$ best.

# 3 Discussion

In conclusion, Models 2 and 2c have been most successful in reproducing the velocity characteristics of the scattered upstream light in HH 1. They differ most significantly in their implied values for the magnitude of the relative velocity between the source and the scatterers $u_0$. Model 2 predicts $u_0 = 270\,\mathrm{km\,s^{-1}}$, which, although still somewhat higher than the $\sim 200\,\mathrm{km\,s^{-1}}$ found by fitting bow shock models to the line ratios and line profiles of HH 1 (Raga et al. 1988), is lower than that derived from proper motion measurements. Hence, if the proper motion values are taken at face value to represent the true speed of the shock-excited gas in HH 1 (rather than merely a pattern speed), then Model 2 provides *some* support for the notion that the bow shock of HH 1 is propagating into a moving environment. Such a situation could arise as the result of a variable velocity jet, as discussed in the introduction. Model 2c, however, lends no support to such a notion since it predicts a $u_0$ similar to the proper motion. Although current observational evidence makes it difficult to discriminate between these two models, spectropolarimetric observations of the scattered light could help considerably. This is because, as can be seen from the figures, the peak velocity of the *polarized intensity* is roughly constant at $v_{\mathrm{peak}} \sim 0.9 u_0$ for *all* upstream displacements and for *all* the models. Measurement of this quantity would hence provide a relatively unambiguous determination of $u_0$. The one factor that might spoil this rosy picture is the uncertainty over the source line profile (as seen by the scatterers). All the above models were calculated using a Gaussian profile for the source with a dimensionless velocity width $w = 0.1$, corresponding to a FWHM of $45\,\mathrm{km\,s^{-1}}$ for Model 2, or $55\,\mathrm{km\,s^{-1}}$ for Model 2c. This is slightly lower than the observed value for the source ($79\,\mathrm{km\,s^{-1}}$) but theoretical models of bow shock spectra (Raga 1985) show that, although, for a bow shock in the plane of the sky, many emission lines have velocity widths comparable with the velocity of the bow shock, the upstream dust would see a much narrower line profile, albeit with broad faint wings. Figure 9 shows the result of Model 2e, which is identical to Model 2 but with $w = 0.2$. For this model, the best-fit value of $u_0$ is $230\,\mathrm{km\,s^{-1}}$, which is much more in keeping with the bow shock models and the moving medium hypothesis. The RMS deviation of the fit from the observed velocity peaks is, at $17\,\mathrm{km\,s^{-1}}$, only marginally worse than Model 2 and the fit to $\Delta v_{\mathrm{FWHM}}$ is considerably better. The polarized intensity $v_{\mathrm{peak}}$ is at $\sim 0.85 u_0$ for this model, so it can be seen that if the source velocity width is uncertain by a factor two, there will be an uncertainty in the derivation of $u_0$ by this method of $\lesssim 10\%$.

Although the results presented here seem to rule out a Rayleigh form for the scattering phase function, it is quite possible that such a model *may* be capable of being fitted to the observations by, for example, adopting a density distribution different from the homogeneous one assumed here or by introducing an internal velocity field among the scatterers. However, such further complications would necessarily be rather ad hoc although their introduction may become necessary, once spectropolarimetric observations are obtained, if the simple models developed so far no longer suffice.

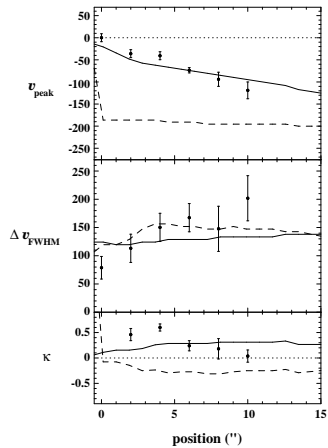

Figure 9: The same as Figure 4 but for Model 2e, which differs from Model 2 in having a source line profile with double the dimensionless velocity width. The source velocity magnitude $u_0 = 230\,\mathrm{km\,s^{-1}}$. The filled circles show the observational results (total intensity) of SB.

To summarize, the results of this case study demonstrate the powerful potential of emission line spectropolarimetry for studying astrophysical flows. The relatively simple geometry of HH 1 allows the polarized scattered light to reveal an "image" of the bowshock at the head of the jet, as seen from the upstream dust. The Doppler shift of this light is then a *direct* measure of the velocity of the bowshock with respect to the dust. It is therefore of paramount importance to conduct spectropolarimetric observations of this object in order to constrain the various possible models.

WJH gratefully acknowledges financial support from SERC, UK and CONACyT, México during the course of this work.